\documentclass[12pt]{article}          % \documentclass[10pt,twocolumn]{article}
\usepackage{graphicx,epsf,subfigure,pstricks,pst-node,psfrag,amsthm,amssymb,amsmath,url,inputenc}
\begin{document}

\title{\textbf{\huge Nonlinear Generalized Ridge Regression}}
\author{Robert L. Obenchain \\ Risk Benefit Statistics}
\date{October 2023}
\maketitle

\begin{abstract}
\noindent A Two-Stage approach is described that literally ``straightens out'' any potentially nonlinear relationship
between a $y-$outcome variable and each of $p \ge 2$ potential $x_i-$predictor variables. The $y-$outcome is then
predicted from all $p$ of these ``linearized''  $s(x_i)-$predictors using the form of Generalized Ridge Regression that
is \textit{most likely} to yield minimal MSE risk under Normal distribution-theory. These estimates are then compared
and contrasted with those from the Generalized Additive Model that uses the same $x_i$ variables.

\end{abstract}

\section{Introduction} % Section 1...
We discuss a simple ``Two-Stage'' approach that combines use of nonlinear \textit{gam()} smooth ``spline'' functions from
the \textit{mgcv} $R-$package, Wood (2001,2023), Wood and Augustin (2002), Wood, Pya and Saefken (2016) in stage one, with optimally
biased estimation of linear models using the \textit{eff.ridge()} function from the \textit{RXshrink} $R-$package,Obenchain (2022,2023),
in stage two.

This new multiple regression strategy is applicable when $p \ge 2$ different individual $x-$vectors are initially used
alone (one-at-a-time) to make $p$ (different) nonlinear predictions of a single observed $y-$outcome vector. Especially when the number,
$n$, of observations available is large [e.g. $n = 2,793$ in the example analyses], using the gam() function from the \textit{mgcv} $R-$package,
Wood (2023), provides superior speed and efficiency.

\begin{table}[ht]
\begin{tabular}{@{}ccllrr@{}}
\multicolumn{5}{c} {\textbf{TABLE 1 -- Names and Descriptions of Variables}} \\
Var & Name & Description & Min & Max \\
y  & AACRmort & Age Adjusted Circulatory-Respiratory Mortality & 95.5 & 46.00 \\
x1 & Avoc     & Anthropogenic Volatile Organic Compounds & 0.22282 & 2.8906 \\
x2 & Bvoc     & Biogenic Volatile Organic Compounds & 0.2611 & 3.3092 \\
x3 & PREMdeath & Premature Death Index (before Age 75) & 2853 & 36469 \\
x4 & ASmoke  & Adult Smoking Percentage & 0.06735 & 0.41202 \\
x5 & ChildPOV & Children Living in Poverty Index & 0.0290 & 0.6630 \\
x6 & IncomIEQ & Income Inequality Index & 2.932 & 8.929 \\
\end{tabular}
\end{table}

Our numerical example will use $p = 6$ $x-$predictors of $y-$Outcome values of ``Age Adjusted Circulatory-Respiratory Mortality'' rates
for $2,793$ US Counties in $2016$. A CSV file containing the data can be downloaded from \textit{Dryad}, Young and Obenchain (2022).

Upon completing this ``Initial Phase'' of analysis, the analyst forms an $R-$data.frame containing the following $2*p + 1$ variables:
$(a)$ the observed $y-$outcome vector, $(b)$ the $p$ $x-$predictor variables, and $(c)$ the $p$ $y-$predictor vectors, each of which
is \textit{a semi-parametric, nonlinear function} of only one of the given $x-$predictor vectors. Note that $gam()$ fitting uses a form of
Generalized Ridge Regression that minimizes a Generalized Cross Validation (GCV) statistic to provide ``automatic smoothing''.

Finally, the analyst uses the eff.ridge() and MLcalc() functions from the ``RXshrink'' $R-$package, Obenchain (2023), to first generate
``TRACE'' graphics such as those displayed in Figure $4$ and then to calculate the relative $MSE$ risk statistics reported in Table $3$.
Again, we find that the spline estimates with $Maximum Likelihood$ of being ``optimally biased'' under Normal distribution-theory,
Obenchain (2022), generally tend to achieve both lower ``residual error'' and lower ``relative'' MSE risk compared to \textit{unbiased}
(BLUE) least squares estimates.

\section{Linear Rank Deficiency} % Section 2...

Regression models typically assume that (1) the n by p ``centered'' $X-$matrix, $(I-11'/n)X$, is of
full (column) rank, $p$, and that (2) an ``intercept'' term, $\mu$, is included to assure that the ``fit''
(a line, plane or hyper-plane) passes through $y = \bar{y}$ in the $p+1$ dimensional Euclidean space where
all $x-$coordinates equal their mean vector. Specifically,

\begin{equation}
\hat{\mu} = \bar{y} - \bar{x}'\hat{\beta}\text{ .}  \label{IIT}
\end{equation}
  
\noindent Note that this \textit{intercept estimate} automatically changes when the $\hat{\beta}$ coefficient vector
is shrunken via generalized ridge regression, allowing the implied ``fit'' to pivot about $\bar{x}$ in $p-$space.

\section{Visual Validation using Spline $x-$Coordinates} % Section 3...

The \textit{syxi()} and \textit{plot.syxi()} functions in Version $2.3$ of the ``RXshrink'' CRAN $R-$Package, Obenchain (2023),
compute and display three types of predictions of a single $y-$outcome variable. The two \textit{scatter plots} in each row of Figures $1$,
$2$ and $3$ display $x_i$ vs. $y$ in the left-hand plot and the $gam()$ spline $s(x_i)$ fitted values vs. $y$ in the right-hand plot. Note
that each left-hand scatter plot contains both a purely ``linear'' $lm(y \sim x_i)$ fit (dashed RED) and a clearly ``nonlinear'' fit (BLUE).

The right-hand plots in these Figures provide the statistical information that an analyst needs to perform ``Visual Validation.''
Specifically, the fitted spline \textit{abscissa} in each right-hand scatter provides a nonlinear transformation that, in my experience,
frequently tends to make the  $lm(y \sim s(x_i))$ fit (solid BLUE) ``LOOK'' rather reasonable. On the other hand, this transformed scatter may
lack homogeneity of variance.  % homoscedasticity

\begin{figure}
\center{\includegraphics[width=6in]{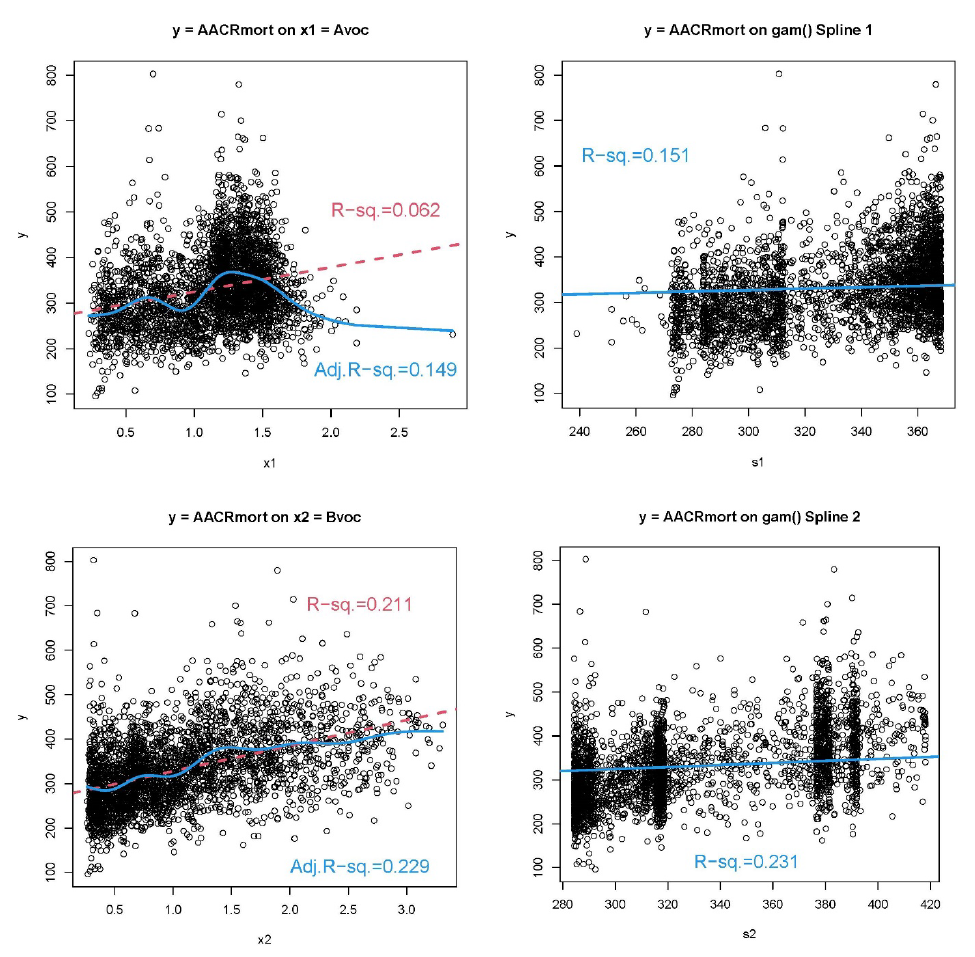}}
\caption{\label{fig:GAM12} The initial pair of $gam()$ predictions of \textit{AACRmort} $y-$outcomes are the $x1 = Avoc$ and $x2 = Bvoc$
Volatile Organic Components of $PM2.5$ air pollution. Note that the relationship between $Avoc$ and $AACRmort$ is distinctly
non-monotone, while $Bvoc$ has nearly monotone (possibly causal) effects at its higher levels. While $Avoc$ and $x6 = IncomIEQ$ are the
two least important predictors of mortality, $Bvoc$ is slightly more important.}
\end{figure} 

\begin{figure}
\center{\includegraphics[width=6in]{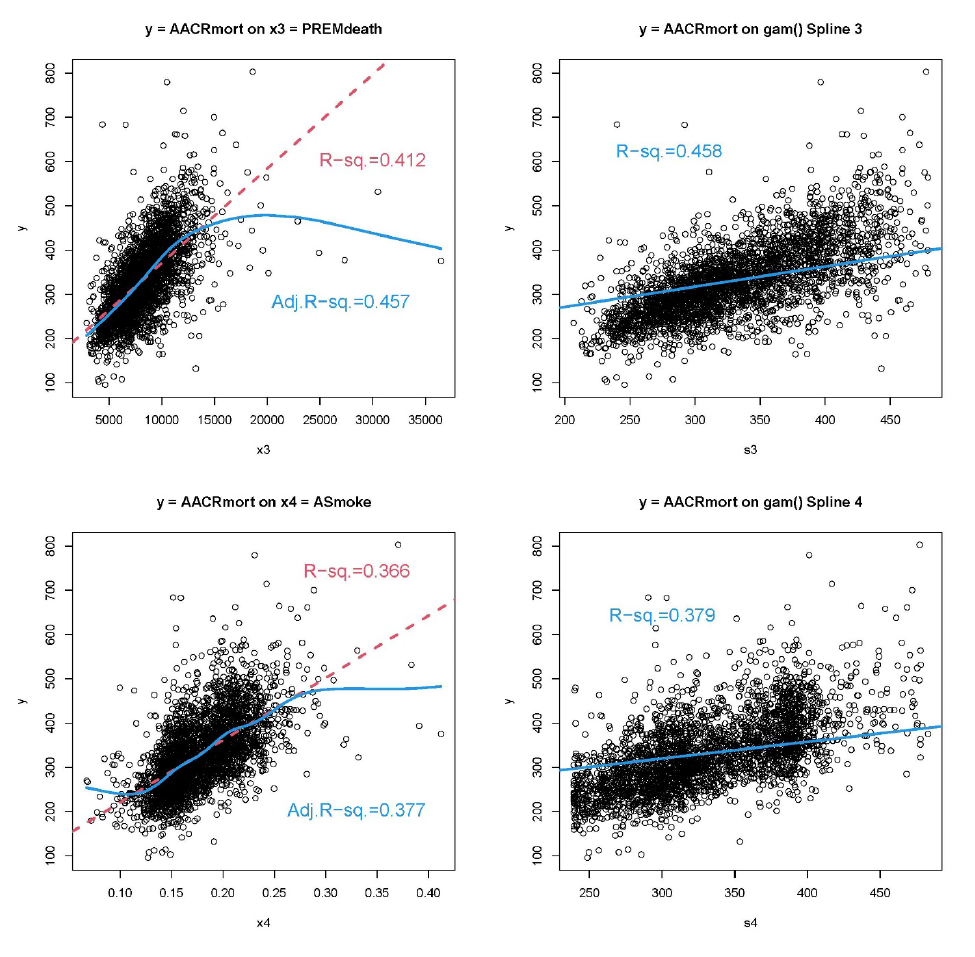}}
\caption{\label{fig:GAM34} The third and fourth pairs of $gam()$ predictions of \textit{AACRmort} $y-$outcomes are shown here. $x3 = PREMdeath$ is quite curved and has the highest Adjusted R-square, confirmed by $s3$. $x4 = ASmoke$ has nearly monotone effects on mortality and the second
highest Adjusted R-square, confirmed by $s4$.}
\end{figure}

\begin{figure}
\center{\includegraphics[width=6in]{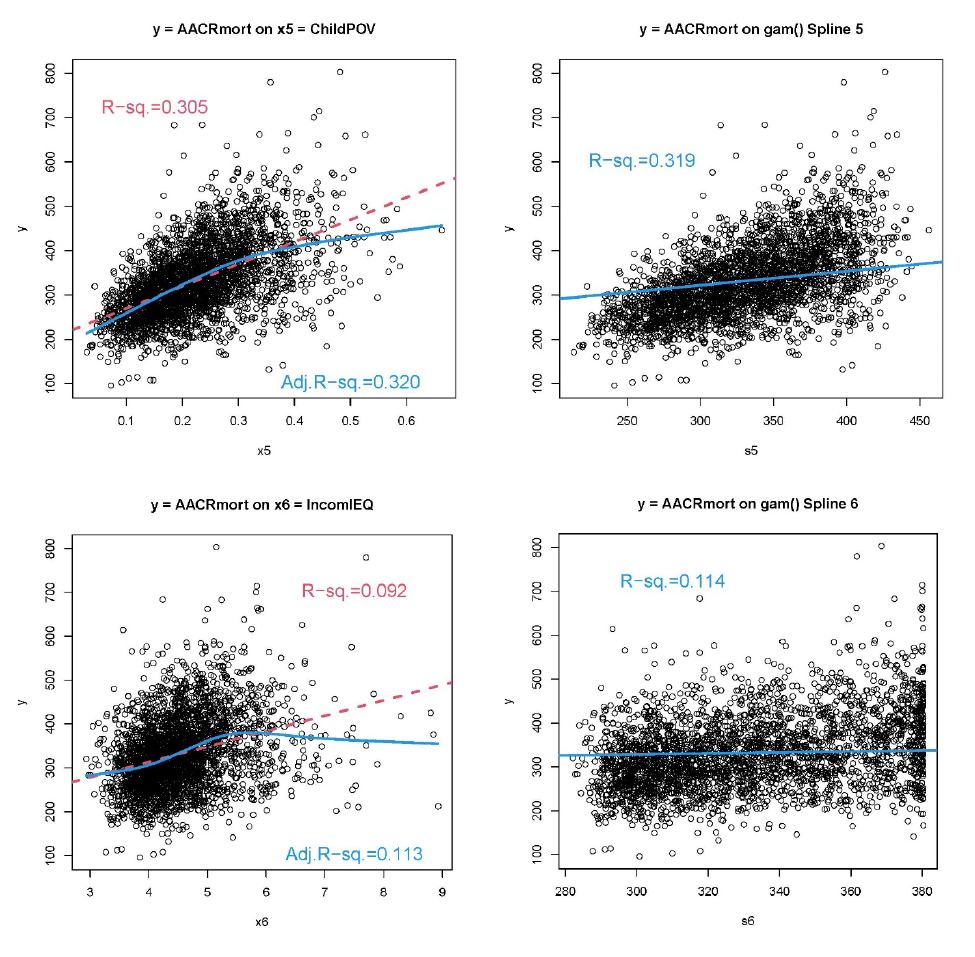}}
\caption{\label{fig:GAM56} The fifth and sixth pairs of $gam()$ predictions of $y-$outcomes are ($x5 = ChildPOV$, $s5$) and ($x6 = IncomIEQ$,
$s6$). While $x5 = ChildPOV$ is the third most important predictor of mortality, $x6 = IncomIEQ$ is not very important.}
\end{figure}

\section{Linear GRR Shrinkage} % Section 4...

Early proposals for choosing the ``k-factor'' defining the $1-$parameter shrinkage path of Hoerl
and Kennard (1970) were based mostly upon ``heuristics.'' In sharp contrast, shrinkage
based upon \textit{Normal-theory Maximum Likelihood estimation} has a firm theoretical foundation,
Obenchain (1975, 1977, 1978). The \textit{Efficient} $p-$parameter shrinkage-path is a
\textit{two-piece linear spline} with a \textit{single interior Knot} at the overall ``best''
(minimum MSE risk) vector of point-estimates, Obenchain (2022).
 
Ridge computations and \textit{TRACE} displays for the Efficient Path (plus other visualizations)
are implemented in version $2.3$ of the \textit{RXshrink} $R-$package, Obenchain (2023). Each \textit{TRACE}
typically displays estimates of $p \geq 2$ quantities that change as \textit{shrinkage} occurs.
The ``coef'' \textit{TRACE} displays the $p$ fitted linear-model $\beta-$coefficients, while the ``rmse''
\textit{TRACE} plots \textit{relative} mean-squared-error estimates given by the diagonal elements of the
MSE-matrix divided by the OLS-estimate of $\sigma^2$.

\section{Overall Extent of Shrinkage} % Section 4...

The \textit{multicollinearity allowance}, $m$, measures the ``extent'' of shrinkage
applied.
\begin{equation}
m=p-\delta_1-\cdots-\delta_p=rank(X)-trace(\Delta).  \label{MCAL}
\end{equation}

\noindent Note that $0 \leq m \leq p$, Obenchain (1977). Besides being the rank of $X$, $p$ is also the
$trace$ of the OLS Hat-matrix, $XX^+$. Similarly, $trace(\Delta)$ is like a measure of ``rank''
for the \textit{diagonal} ($p \times p$) $\Delta$ shrinkage-factor matrix.

\section{Correlations within Groups of Variables} 

Note that the $2,793$ observed $y-$Outcomes are positively correlated with the six given $x-$variables and
also tend to be slightly more positively correlated with the six $gam()$ spline prediction vectors, as shown in
Table $2$. 
 
\begin{table}[ht]
\begin{tabular}{@{}lccccccccl@{}}
\multicolumn{10}{c} {\textbf{TABLE 2 -- Pearson Correlations within Two Sets of Seven Variables}} \\
variable & y & & x1 & x2 & x3 & x4 & x5 & x6 \\
y   & 1.0000 & & & & & & & \\
x1  & 0.2489 & & 1.00000 & & & & & \\
x2  & 0.4589 & & 0.58472 & 1.0000 & & & & \\
x3  & 0.6421 & & 0.08896 & 0.4217 & 1.00000 & & & \\
x4  & 0.6047 & & 0.32707 & 0.4622 & 0.67611 & 1.0000 & & \\
x5  & 0.5524 & & 0.11336 & 0.4884 & 0.69932 & 0.6605 & 1.0000 & \\
x6  & 0.3040 & & 0.13933 & 0.4163 & 0.41804 & 0.3800 & 0.5708 & 1.0000 \\
    &        & &         &        &         & & & \\
variable & y & & s1 & s2 & s3 & s4 & s5 & s6 \\
y   & 1.0000 & & & & & & & \\        
s1 & 0.3888 & & 1.0000 & & & & & \\
s2 & 0.4809 & & 0.5991 & 1.0000 & & & & \\
s3 & 0.6769 & & 0.3077 & 0.5117 & 1.0000 & & & \\
s4 & 0.6156 & & 0.5002 & 0.5485 & 0.6621 & 1.0000 & & \\
s5 & 0.5656 & & 0.3029 & 0.5192 & 0.7348 & 0.6607 & 1.0000 & \\
s6 & 0.3379 & & 0.1884 & 0.4378 & 0.4389 & 0.3775 & 0.5775 & 1.0000 \\    
\end{tabular}
\end{table}

\section{MSE Risk Comparisons}

Applied researchers should consider transforming the individual $x-$variables available to them using nonlinear
spline methods to potentially improve prediction of their target $y-$outcome variable.
The \textit{one-variable-at-a-time} first stage approach illustrated here provides a viable way to start. Once a researcher
identifies $p \ge 2$ predictor variables, two different sets of Maximum Likelihood estimates under Normal distribution-theory
(Unbiased or Optimally Shrunken) can be computed and compared in several ways.

\begin{table}[ht]
\begin{tabular}{@{}lccccccl@{}}
\multicolumn{8}{c} {\textbf{TABLE 3 -- Key Statistical Comparisons for the Linear and NL Models}} \\
                     &             &        &        &        &        &        \\
Linear Model Formula & $y \sim x1$ & $+ x2$ & $+ x3$ & $+ x4$ & $+ x5$ & $+ x6$ \\
Residual Mean Square & 0.506395 & & & & & \\
Residual Std. Error  & 0.711615 & & & & & \\
OLS Beta Coefficients     & 0.060975 & 0.136652 & 0.389470 & 0.216750 & 0.099290 & -0.063270 \\ 
ML Optimally Biased Betas & 0.064848 & 0.134431 & 0.384555 & 0.218586 & 0.101132 & -0.062424 \\
OLS Relative MSE Risks    & 0.000613 & 0.000745 & 0.000821 & 0.000822 & 0.000968 &  0.000519 \\
ML Minimum Relative Risks & 0.000292 & 0.000428 & 0.000756 & 0.000380 & 0.000524 &  0.000488 \\
dMSE Estimates            & 0.9996   & 0.9413   & 0.9964   & 0.9429   & 0.9728   &  0.003282 \\
                     &             &        &        &        &        &       \\
Transformed Model Formula & $y \sim s1$ & $+ s2$ & $+ s3$ & $+ s4$ & $+ s5$ & $+ s6$ \\
Residual Mean Square & 0.479143 & & & & & \\ 
Residual Std. Error  & 0.692202 & & & & & \\
OLS Beta Coefficients     & 0.098704 & 0.056765 & 0.451541 & 0.214582 & 0.035727 & -0.005387 \\
ML Optimally Biased Betas & 0.103079 & 0.056651 & 0.443230 & 0.221146 & 0.034597 & -0.006164 \\
OLS Relative MSE Risks    & 0.000590 & 0.000715 & 0.000852 & 0.000815 & 0.000996 &  0.000533 \\
ML Minimum Relative Risks & 0.000380 & 0.000559 & 0.000911 & 0.000790 & 0.000939 &  0.000453 \\  
dMSE Estimates            & 0.9997   & 0.4332   & 0.9961   & 0.8721   & 0.9359   &  0.9882 \\
\end{tabular}
\end{table}

\begin{figure}
\center{\includegraphics[width=6in]{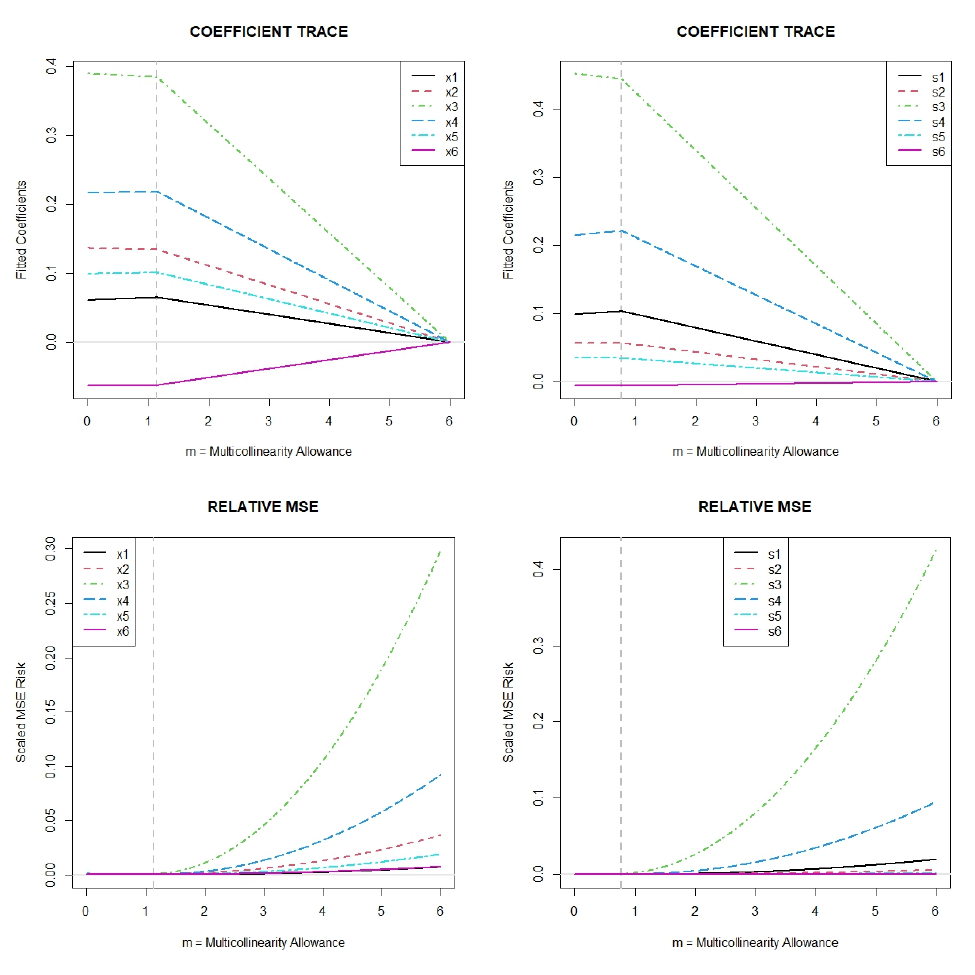}}
\caption{\label{fig:TR4} The two Traces of $\beta-$coefficients in the Top Row show that relative
magnitudes of coefficient estimates are slightly unstable, with less shrinkage (to $m = 0.775$ rather than
to $m = 1.07$) being appropriate when all six nonlinear transformations of the given $x-$variables
are used in the model. The corresponding pair of \textit{Relative Risk} ($MSE/\sigma^2$) Traces in the bottom row
confirm that the six nonlinear predictors of mortality (AACRmort) rates provide better predictions. See Table 3
for more detail on this finding.}
\end{figure}

Note that Table $3$ (on page $9$) shows that the shrunken $gam()$ model achieves not only lower residual error than
the shrunken Linear model but also lower MSE risks ``relative to'' that lower residual error for $5$ of the $6$
spline predictor variables!

Note that the optimally shrunken $x3 = PREMdeath$ coefficient has an estimated MSE risk of $0.5064 * 0.000756 = 0.000383$,
while that of its $s3$ counterpart is estimated to be larger: $0.4791 * 0.000911 = 0.000436$.
All differences in MSE risks and relative risks again tend to be small in this large scale EPA Particulate Matter
example, but achieving numerical risk reductions for $5$ out of $6$ relevant predictor variables certainly suggests
that this simple ``Two-Stage'' approach deserves further study of its ability to estimate the effects of ``x'' and
``s'' variables and to predict ``y-outcomes''.

\section{Summary}

Here, we have provided a fairly large example ($p = 6$ $x-$predictors and $n = 2,793$ observations) of our ``two-stage''
approach. This appears to be a straight-forward ``common sense'' way to transform predictor variables and, ultimately, to
end up with potentially ``better'' unbiased and/or shrunken  predictions of the $y-$outcome.

In early arXiv papers on this topic, we called this two-stage approach ``non-parametric''. Others might prefer to
call it ``semi-parametric smoothing''. Any transformation applied to an $x-$variable in stage-one \textit{must be nonlinear}! 
  
Software implementing accurate computations and providing clear ``visual insights'' into the strengths and weaknesses
of alternative methods are indispensable components of an adequate \textit{Tool Bag} for today's applied researchers.

\section{References}

\noindent Hastie, T. and Tibshirani, R. (1990). ``Generalized Additive Models.'' \textit{Chapman and Hall}.\\

\noindent Obenchain, R. L. (1975). ``Ridge analysis following a
preliminary test of the shrunken hypothesis.'' \textit{Technometrics} 17,
431$-$441. \url{http://doi.org/10.1080/00401706.1975.10489369}\\

\noindent Obenchain, R. L. (1977). ``Classical F-tests and confidence regions
for ridge regression.'' \textit{Technometrics} 19, 429$-$439.
\url{http://doi.org/10.1080/00401706.1977.10489582}\\

\noindent Obenchain, R. L. (1978). ``Good and Optimal Ridge Estimators.''
\textit{Annals of Statistics} 6, 1111$-$1121. \url{http://doi.org/10.1214/aos/1176344314}\\

\noindent Obenchain, R. L. (2022). ``Efficient Generalized Ridge Regression'',
\textit{Open Statistics} 3, 1$–$18. \url{https://doi.org/10.1515/stat-2022-0108}\\

\noindent Obenchain, R. L. (2023). ``\textit{RXshrink}: Maximum Likelihood
Shrinkage using Generalized Ridge or Least Angle Regression Methods'', ver 2.3,
\url{https://CRAN.R-project.org/package=RXshrink}\\

\noindent Wood, S. N. (2001), ``mgcv: GAMs and Generalized Ridge Regression for R.'' \textit{R News}
1(2) :20$-$25.\\

\noindent Wood, S. N. and Augustin, N. H. (2002), ``GAMs with integrated model selection using
penalized regression splines and applications to environmental modelling.'' \textit{Ecological
Modelling} 157: 157$-$177.\\

\noindent Wood, S. N., Pya N. and Saefken B. (2016), ``Smoothing parameter and model selection for general
smooth models (with discussion). \textit{Journal of the American Statistical Association} 111: 1548$-$1575.
\url{http://doi:10.1080/01621459.2016.1180986}\\

\noindent Wood, S. N. (2023), ``mgcv: Mixed GAM Computation Vehicle with Automatic Smoothness
Estimation'', ver. 1.8-42, \url{https://CRAN.R-project.org/package=mgcv}\\

\noindent Young, S. S. and Obenchain, R. L. (2022). ``EPA Particulate Matter Data'', Dryad [Data Archive].
\url{https://doi.org/10.5061/dryad.63xsj3v58}\\

\end{document}